\documentclass[PRL,preprintnumbers,floatfix,aps,notitlepage,nofootinbib,twocolumn,showpacs,amssymb]{revtex4-2}
\usepackage{amsmath,amsfonts,bm}
\usepackage{graphicx}
\usepackage{cancel}
\usepackage[colorlinks, linkcolor={red},citecolor={blue}]{hyperref}

\begin{document}

\title{Warped vacuum energy by black holes}
\author{J. Ovalle$^{ab}$}
\email{jorge.ovalle@physics.slu.cz}
\affiliation{$^a$Research Centre for Theoretical Physics and Astrophysics,
	Institute of Physics, Silesian University in Opava, CZ-746 01 Opava, Czech Republic.\\ \\
$^b$Departamento de F\'isica, Facultad de Ciencias B\'asicas,
Universidad de Antofagasta, Chile.}

\begin{abstract}
	The belief still persists, at least for an important part of the community, that in general relativity the cosmological constant $\Lambda$  must remain immaculate. However, Bianchi identities show that $\Lambda$ could acquire local properties in the presence of matter, i.e., $\Lambda\rightarrow\,\Lambda(x^\mu)$. In this paper we go even further, and we show that a non-uniform vacuum energy does not even require coexistence with any additional form of matter-energy, nor of any theory beyond general relativity, at least for extreme environments, such as the proximity of a black hole.
		\end{abstract} 
\maketitle
%
%
%
\section{Introduction}
There is consensus that one of the greatest challenges for theoretical physics is to explain the cosmological constant $\Lambda$~\cite{Weinberg:1988cp,Martin:2012bt}. Since its introduction by Einstein in 1917, it has had a rather controversial history, including its withdrawal and again its inclusion with different interpretations. The story changes radically after 1998, when $\Lambda$ is promoted as the ideal source that generates the accelerated expansion of the universe, that is, the so-called dark energy~\cite{SupernovaSearchTeam:1998fmf,SupernovaCosmologyProject:1998vns,Peebles:2002gy,Huterer:2017buf}. Concurrently, the interpretation of $\Lambda$ as the vacuum energy arises naturally from Einstein's equations, which leads to identify the energy-momentum tensor for the vacuum as $\kappa\,{T}^{(\rm vac)}_{\mu\nu}={\Lambda}\,g_{\mu\nu}$, and the reason the concepts of cosmological constant, vacuum energy and dark energy are usually mentioned interchangeably. 

Regarding the vacuum state and its tensorial representation ${T}^{(\rm vac)}_{\mu\nu}$, it has been established, as a consequence of Bianchi identities, that this vacuum state is uniform in energy density and isotropic in
pressure. This is a quite strong result, and the consensus assures that the only way to bypass this statement is by considering theories beyond general relativity, or a vacuum state interacting with matter fields, i.e., in those regions where $T_{\mu\nu}\neq\,0$.
The goal of this letter is to show that the symmetry of spacetime plays a fundamental role in bypassing the established framework for vacuum energy, namely,  its uniformity and isotropy in the context of general relativity. Contrary to common belief, we will demonstrate the existence of a cosmological constant or vacuum energy with local properties, without filling the $\Lambda$-vacuum with any additional form of matter-energy, or using any alternative theory to general relativity. This is achieved by a continuous deformation of the vacuum state ${T}^{(\rm vac)}_{\mu\nu}$, carried out in full agreement with Bianchi identities.
This deformation, which yields a non-uniform and anisotropic vacuum state, occurs mainly in the vicinity of rotating systems. However, it may produce large scale consequences, and could give some insight to understand gravitational phenomena without conjecturing exotic forms of matter. To carry out the above, we use the recent revisited Kerr-de Sitter black hole solution~\cite{Ovalle:2021jzf}, whose main characteristic is the appearance of a warped curvature due to rotational effects, which we will explain in details in terms of a warped vacuum energy.


\section{Cosmological constant and vacuum energy}
\label{sec2}
In order to highlight the fact that the cosmological constant $\Lambda$ may acquire local properties, we briefly review the interpretation of $\Lambda$ in terms of the vacuum energy. Let us start with Einstein equations~\footnote{Our signature is $-2$ and we use units with $c=1$
	and $\kappa=8\,\pi\,G_{\rm N}$, where $G_{\rm N}$ is Newton's constant.} 
\begin{equation}
	\label{ee}
	R_{\mu\nu}-\frac{1}{2}\, R\,  g_{\mu\nu}-\Lambda\,  g_{\mu\nu}
	=
	\kappa\,{T}_{\mu\nu}\ ,
\end{equation}
where we can identify the energy-momentum tensor for the vacuum as
\begin{equation}
	\label{vac}
	{T}^{(\rm vac)}_{\mu\nu}
	=\frac{\Lambda}{\kappa}\,g_{\mu\nu}\ ,
\end{equation}
which describes a vacuum state defined in a spacetime with curvature $R=-4\Lambda$, whose energy density and isotropic pressure are given by $\epsilon^{(\rm vac)}=-p^{(\rm vac)}=\frac{\Lambda}{\kappa}$. Since  the  Einstein  tensor  in  Eq.~\eqref{ee} satisfies the contracted Bianchi identities  $\nabla_\mu\,G^{\mu}_{\,\,\nu}=0$, we have
\begin{equation}
	\label{exchange2}
	\partial_\nu\Lambda=-\kappa\nabla_\mu\,T^{\mu}_{\,\,\,\nu}\ .
\end{equation}
Let us focus now on expression~\eqref{exchange2}, which is very significant. First of all, we see that for those regions where $T^{\mu}_{\,\,\,\nu}\neq\,0$, we have two possible options. One is that $\Lambda$ is constant everywhere and therefore $T^{\mu}_{\,\,\,\nu}$ is covariantly conserved, i.e., $\nabla_\mu\,T^{\mu}_{\,\,\,\nu}=0$. The second option, which requires a local cosmological constant $\Lambda\rightarrow\Lambda(x^\mu)$, shows an exchange of energy-momentum between matter and the vacuum energy, as we see through the expression in Eq.~\eqref{exchange2}. As we see, despite the belief of many, general relativity does not prohibit the existence of a non-uniform cosmological constant.

Contrary to the above, an always constant $\Lambda$ seems the only option in the vacuum $T^{\mu}_{\,\,\,\nu}=0$. Therefore, in the absence of matter, the vacuum energy in Eq.~\eqref{vac} is defined in a constant curvature spacetime, which is uniform in energy density and isotropic in pressure. Notice that  we have not made any particular assumptions throughout the expressions in Eqs.~\eqref{ee}-\eqref{exchange2}. Therefore, regarding the definition of vacuum energy in the presence of a cosmological constant, we should conclude there is room neither for non-constant curvature, nor for non-uniform density nor for anisotropy. 

Next, we show a simple and direct way to evade the non-existence of an anisotropic and non-uniform vacuum. This is achieved in theories beyond Einstein, which can be described by a modified Einstein-Hilbert action as
\begin{equation}
	\label{ngt}
	S_{\rm G}=S_{\rm EH}+S_{\rm X}=\int\left[\frac{(R-2\Lambda)}{2\kappa}+{\cal L}_{\rm M}+{\cal L}_{\rm X}\right]\sqrt{-g}\,d^4\,x\ ,
\end{equation}
where $R$ is the Ricci scalar, ${\cal L}_{\rm M}$ contains any matter fields appearing in the theory and ${\cal L}_{\rm X}$ the Lagrangian density of a new gravitational sector not described by general relativity. If we keep the definition of vacuum as those regions without any matter field, i.e., ${\cal L}_{\rm M}=0$, we find that the vacuum is now filled not only for the cosmological term, but also for a fluid whose energy-momentum tensor is given by
\begin{equation}
	\label{ngt2}
	\Theta_{\mu\nu}=\frac{2}{\sqrt{-g}}\frac{\delta\,(\sqrt{-g}\,{\cal L}_{\rm X})}{\delta\,g^{\mu\nu}}=2\,\frac{\delta\,{\cal L}_{\rm X}}{\delta\,g^{\mu\nu}}-\,g_{\mu\nu}{\cal L}_{\rm X}\ ,
\end{equation}
in consequence, the energy-momentum tensor for the vacuum in Eq.~\eqref{vac} is modified as
\begin{equation}
	\label{vac2}
	\frac{\Lambda}{\kappa}\,g_{\mu\nu}\rightarrow\,\frac{\Lambda}{\kappa}\,g_{\mu\nu}+\Theta_{\mu\nu}\equiv{T}^{(\rm vac)}_{\mu\nu}\ .
\end{equation}
Depending on the nature of ${\cal L}_{\rm X}$, the vacuum energy in Eq.~\eqref{vac2} could be anisotropic, thus generating a space time with this same property, whose curvature is given by
\begin{equation}
	\label{constant2}
	R=-4\Lambda-\kappa\Theta\ ;\,\,\,\,\,\,\,\,\,\,\Theta=\Theta_\mu^{\,\,\,\mu}\ .
\end{equation}
In addition to the above, $\Lambda$ could be non-uniform. In this case, according to Bianchi identities, there is an interaction between $\Lambda$ and the ${\cal L}_{\rm X}$-sector determined by
\begin{equation}
	\label{exchange3}
	\partial_\nu\Lambda=-\kappa\nabla_\mu\,\Theta^{\mu}_{\,\,\,\nu}\ .
\end{equation}
Notice that the source $\Theta_{\mu\nu}$ filling the $\Lambda$-vacuum could represent some exotic forms of matter, as the conjectured dark matter, a non-$\Lambda$ dark energy or even a combination of these. So far everything seems to indicate that the only way to generate a non-constant vacuum energy is by considering a theory beyond general relativity. However, next we will see that it is possible to generate a non-uniform and anisotropic vacuum, without resorting to any action beyond that of Einstein-Hilbert. 
\section{New Kerr-de Sitter solution}
\label{new}
%

\par
We recently reported in Ref.~\cite{Ovalle:2021jzf} a variant of the well-known Carter's Kerr-de Sitter solution~\cite{Carter:1973rla}~(see also Refs.~\cite{Gibbons:1977mu,Hackmann:2010zz,Akcay:2010vt,Lake:2015xca,Stuchlik:2018qyz,Li:2020drn,Stuchlik:2020rls}), whose main characteristic is the appearance of secondary hairs producing a warped curvature. To carry out this, we generate the rotating version of the Schwarzschild-de Sitter line element following the strategy described in Ref.~\cite{Contreras:2021yxe} (see also Refs.~\cite{Burinskii:2001bq,Dymnikova:2006wn,Smailagic:2010nv,Bambi:2013ufa,Azreg-Ainou:2014nra,Dymnikova:2016nlb}), which leads to
\begin{eqnarray}
	\label{newkerrds}
	ds^{2}
	&=&
	\left[\frac{\Delta_\Lambda-a^2\,\sin^{2}\theta}{\rho^2}\right]
	dt^{2}	-
	\frac{{\rho}^{2}}{{\Delta_\Lambda}}\,dr^{2}
	\nonumber
	\\
	&&
	-
	{\rho}^{2}\,d\theta^{2}
	-
	\frac{{\Sigma_\Lambda}\, \sin^{2}\theta}{{\rho}^{2}}\,d\phi^{2}
	\ ,
	\nonumber
	\\
	&&
	+
	\frac{2\, {a}\sin^{2}\theta}{{\rho}^{2}}\left(r^2+a^2-\Delta_\Lambda\right) 
	\,dt\,d\phi\ ,
\end{eqnarray}
with
\begin{eqnarray}
	&&{\rho}^2
	=
	r^2+{a}^{2}\cos^{2}\theta\ ,
	\label{f0}
	\\
	&&{\Delta_\Lambda}
	 = 
	r^2-2\,M\,r
	+{a}^{2}-\frac{\Lambda}{3}\,r^4\ ,
	\label{f22}
	\\
	&&{\Sigma_\Lambda}
	 = 
	\left(r^{2}+{a}^{2}\right)^{2}
	-{\Delta_\Lambda}\,a^2\sin^{2} \theta\ ,
	\label{f33}
\end{eqnarray}
\begin{figure*}[t]
	\centering
	\includegraphics[width=0.40\textwidth]{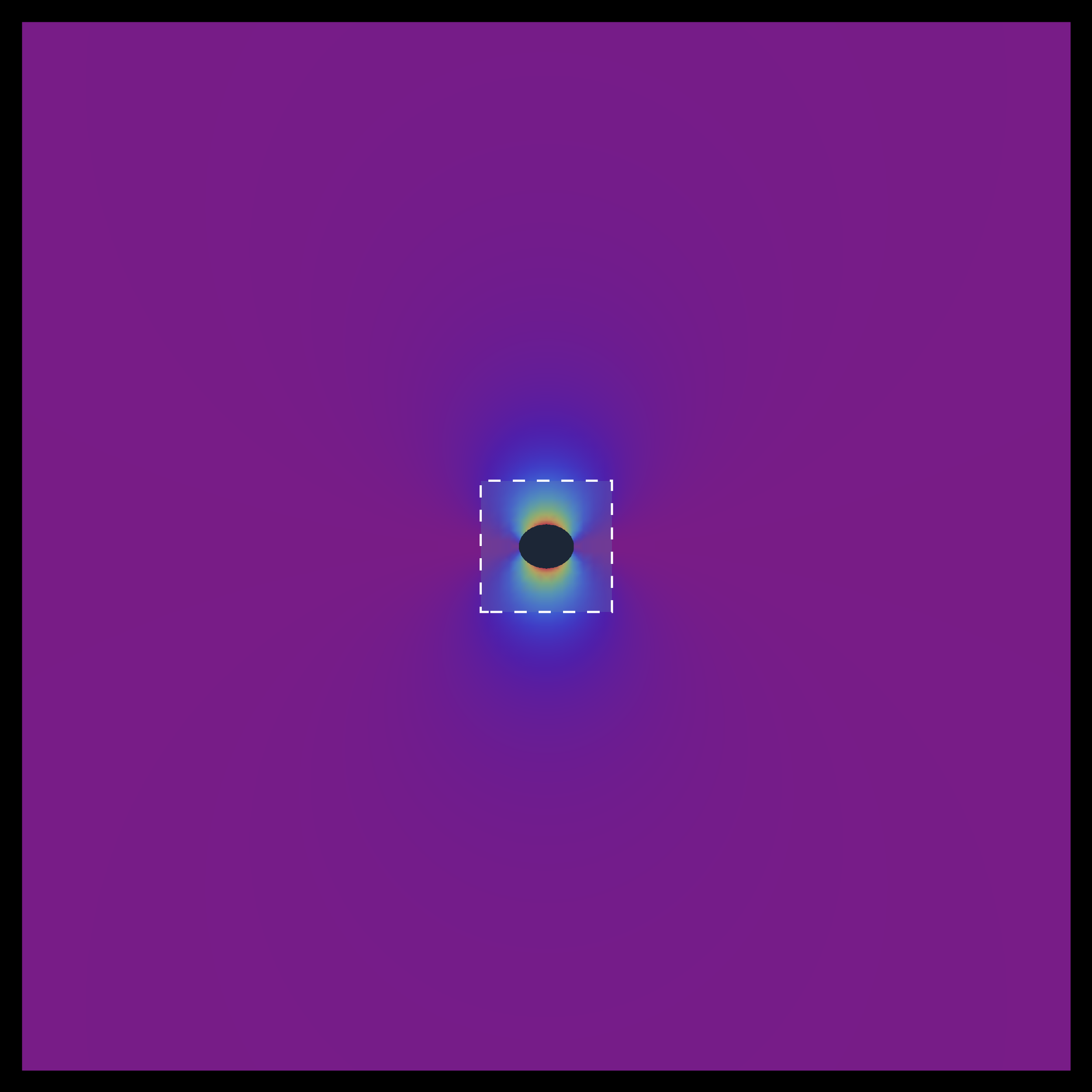} \
	\includegraphics[width=0.045\textwidth]{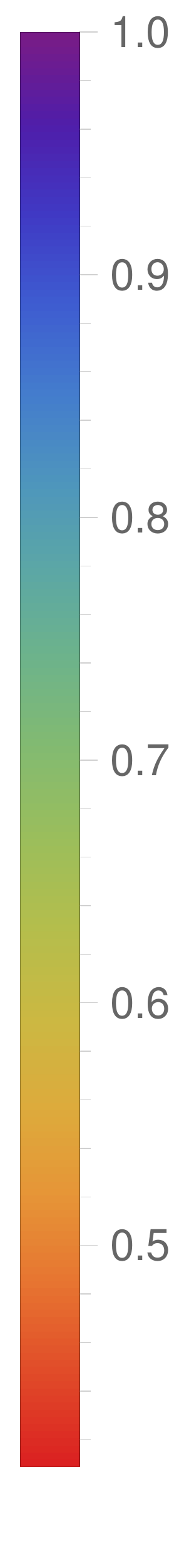} \
	\includegraphics[width=0.40\textwidth]{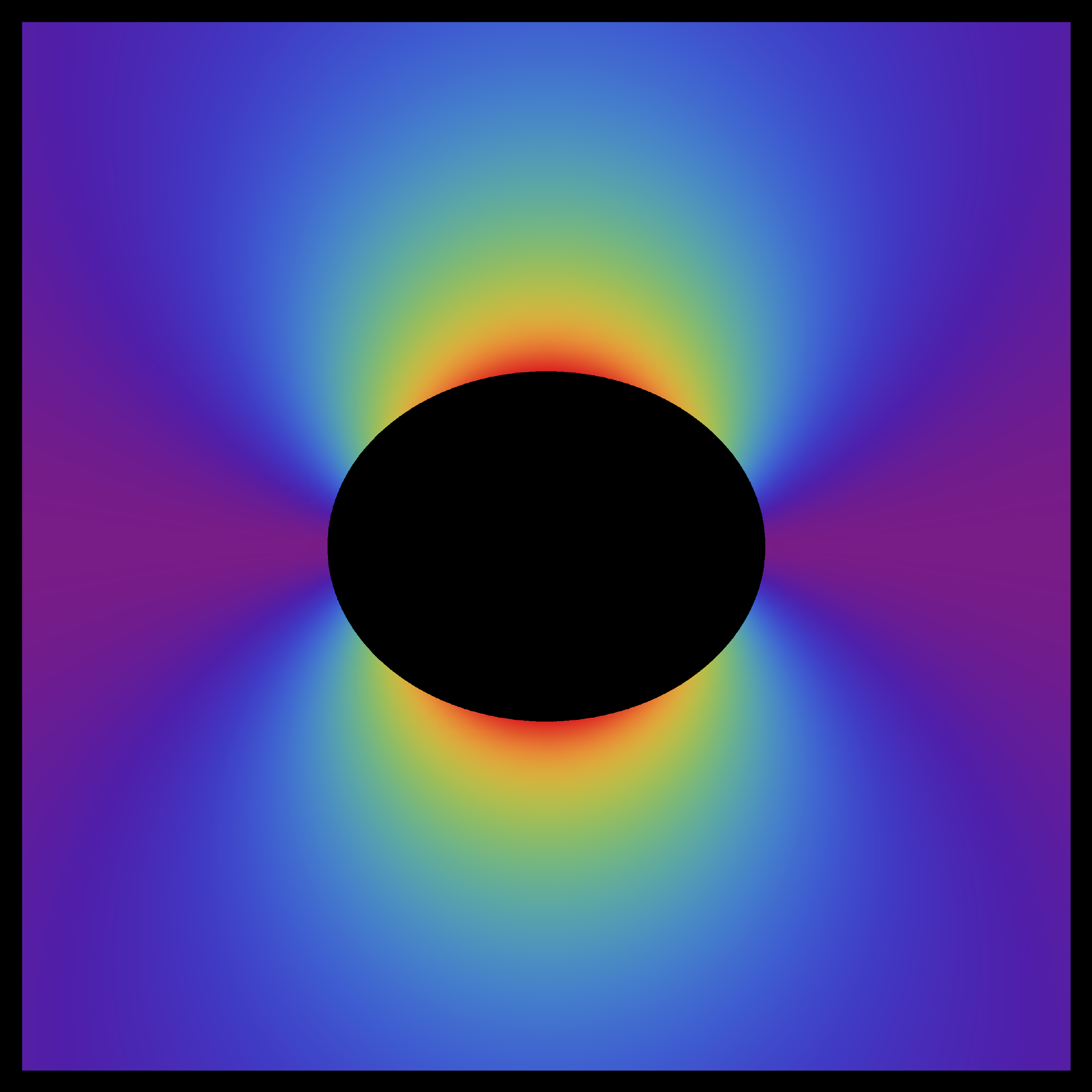} \
	\caption{Warped energy density $\epsilon(r,\theta)$ around black holes. We show the plane $x=0$, where the black region represents a black hole with a vertical rotation axis. There is not warped energy in the equatorial plane, however this is maximum at the poles, in agreement with Eq.~\eqref{energyax}. The plot on the right shows the inner box region magnified. We take $a=0.9$ and ${\Lambda/\kappa}=1$.}
	\label{fig1}
\end{figure*}%

Despite the line element~\eqref{newkerrds} is not a uniform $\Lambda$-vacuum solution, it describes the exterior of a rotating stellar body in a de Sitter or anti-de Sitter background. It was examined in detail in Ref.~\cite{Ovalle:2021jzf}, where we develop a complete and detailed analysis of BHs solutions, precisely identifying the bonds for $M(a,\Lambda)$ within which the existence of BHs is possible. The main feature of this metric is that it shows some  gravitational effects that cannot be elucidated by the conventional solution. One of them consist in the deformation undergone by the curvature $R$ of the spacetime surrounding a rotating stellar distribution, which is given by
\begin{equation}
	\label{R}
	R=-4\,\Lambda\,\frac{r^2}{\rho^2}\neq-4\,\Lambda\ .
\end{equation}
This deformation is particularly significant near the rotating distribution, i.e., 
$r\sim\,a$ and disappears far enough, where $R\sim-4\Lambda$ for $r>>a$. This effect, as expected, is closely related to the warped vacuum energy in the immediate surrounding of extreme gravitational sources, such as black holes. Next we explain this in detail.

First of all, the fact that the line element~\eqref{newkerrds} describes the exterior of a rotating stellar body in a de Sitter or anti-de Sitter background, does not mean a contradiction with the no-hair theorem, since this metric is not a uniform $\Lambda$-vacuum solution, and in consequence secondary hairs appear. In fact, this solution is generated by 
\begin{equation}
	\label{tmunu}
	{T}^{\mu\nu}
	=
	{\epsilon}\, {u}^{\mu}\,{u}^{\nu}
	+{p}_{r}\,{l}^{\mu}\,{l}^{\nu}
	+{p}_{\theta}\,{n}^{\mu}\,{n}^{\nu}
	+{p}_{\phi}\,{m}^{\mu}\,{m}^{\nu}
	\ ,
\end{equation}
where the orthonormal tetrad $\{{u}^{\mu},\,{l}^{\mu},\,{n}^{\mu},\,{m}^{\mu}\}$ reads~\cite{Gurses:1975vu}
\begin{eqnarray}
	{u}^{\mu}
	&=&
	\frac{(r^{2}+{a}^{2})\delta^\mu_0+a\,\delta^\mu_3}{\sqrt{\rho^{2}\Delta}}
	\ ,
	\qquad
	{l}^{\mu}
	=
	\sqrt{\frac{\Delta}{\rho^{2}}}\,\delta^\mu_1
	\label{2}
	\nonumber
	\\
	{n}^{\mu}
	&=&
	\frac{1}{\sqrt{\rho^{2}}}\,\delta^\mu_2
	\ ,
	\qquad
	{m}^{\mu}
	=
	-\frac{{a}\sin^{2}\theta\,\delta^\mu_0+\delta^\mu_3}{\sqrt{\rho^{2}}\sin\theta}
	\ ,
	\label{4}
\end{eqnarray}
and the energy density ${\epsilon}$ and pressures ${p}_r$, ${p}_\theta$, and ${p}_\phi$ satisfy
\begin{eqnarray}
	\label{energyax}
	{\epsilon}
	&=&
	-{p}_{r}
	=
	\frac{\Lambda}{\kappa}\,\frac{r^4}{\rho^4}\ ,
	\\
	\label{pressuresax}
	{p}_{\theta}
	&=&
	{p}_{\phi}
	=
	\epsilon-\frac{2\Lambda}{\kappa}\,\frac{r^2}{\rho^2}
	\ .
\end{eqnarray}
An important rotational effect is the appearance of an anisotropy in the pressures, given by
\begin{equation}
	\label{aniso}
	\Pi\equiv\,p_r-p_\theta=\frac{2\,\Lambda}{\kappa}\,\frac{r^2}{\rho^2}\left(1-\frac{r^2}{\rho^2}\right)\ .
\end{equation}
Also, it is quite easy to check, from Eqs.~\eqref{energyax} and~\eqref{pressuresax}, that the dominant energy condition holds for $\Lambda\,>\,0$, but is violated for $\Lambda\,<0$. The opposite occurs for the strong
energy condition.  Next, we will thoroughly explain how the deformation of the $\Lambda$-vacuum occurs in the vicinity of a rotating object.

\section{Warped vacuum energy}
\label{sec3c}

Let us start by writing the metric~\eqref{newkerrds} in terms of the orthonormal tetrad in Eq.~\eqref{4}, namely,
\label{new2}
\begin{equation}
	g_{\mu\nu}=u_\mu\,u_\nu-l_\mu\,l_\nu-m_\mu\,m_\nu-n_\mu\,n_\nu\ ,
\end{equation}
which plugged into the energy-momentum tensor in Eq.~\eqref{tmunu}, yields 
\begin{equation}
	\label{t1}
	T_{\mu\nu}=\epsilon\,g_{\mu\nu}+(\epsilon+p_\theta)(m_\mu\,m_\nu+n_\mu\,n_\nu)\ .
\end{equation}
We see that the expression in Eq.~\eqref{t1} does not have the form displayed in Eq.~\eqref{vac}. Indeed, it is clear that the energy-momentum tensor was deformed as
\begin{equation}
	\label{t2}
	\frac{\Lambda}{\kappa}\,g_{\mu\nu}\rightarrow\,\epsilon\,g_{\mu\nu}+(\epsilon+p_\theta)(m_\mu\,m_\nu+n_\mu\,n_\nu)
\end{equation}
as a consequence of the black hole rotation. This might tempt us to conclude that the line element~\eqref{newkerrds} does not describe the exterior of a rotating stellar body in a de Sitter or anti-de Sitter background. However, this naive conclusion would be based precisely on the premise we want to avoid, i.e., that the fluid remains immaculate under rotational effects. As we already pointed out, this assumption is too strong, even more so for extreme cases, such as the vicinity of a BH.

Notice that we can write the energy momentum tensor $T_{\mu\nu}$ in Eq.~\eqref{t1} as
\begin{equation}
	\label{t3}
	T_{\mu\nu}=\epsilon\,g_{\mu\nu}+K_{\mu\nu}\ ,
\end{equation}
which clearly contains an isotropic part $\epsilon\,g_{\mu\nu}$ and a tensor
\begin{equation}
	K_{\mu\nu}\equiv\,(\epsilon+p_\theta)\,(m_\mu\,m_\nu+n_\mu\,n_\nu)\ ,
\end{equation}
\begin{figure*}
	\centering
	\includegraphics[width=0.27\textwidth]{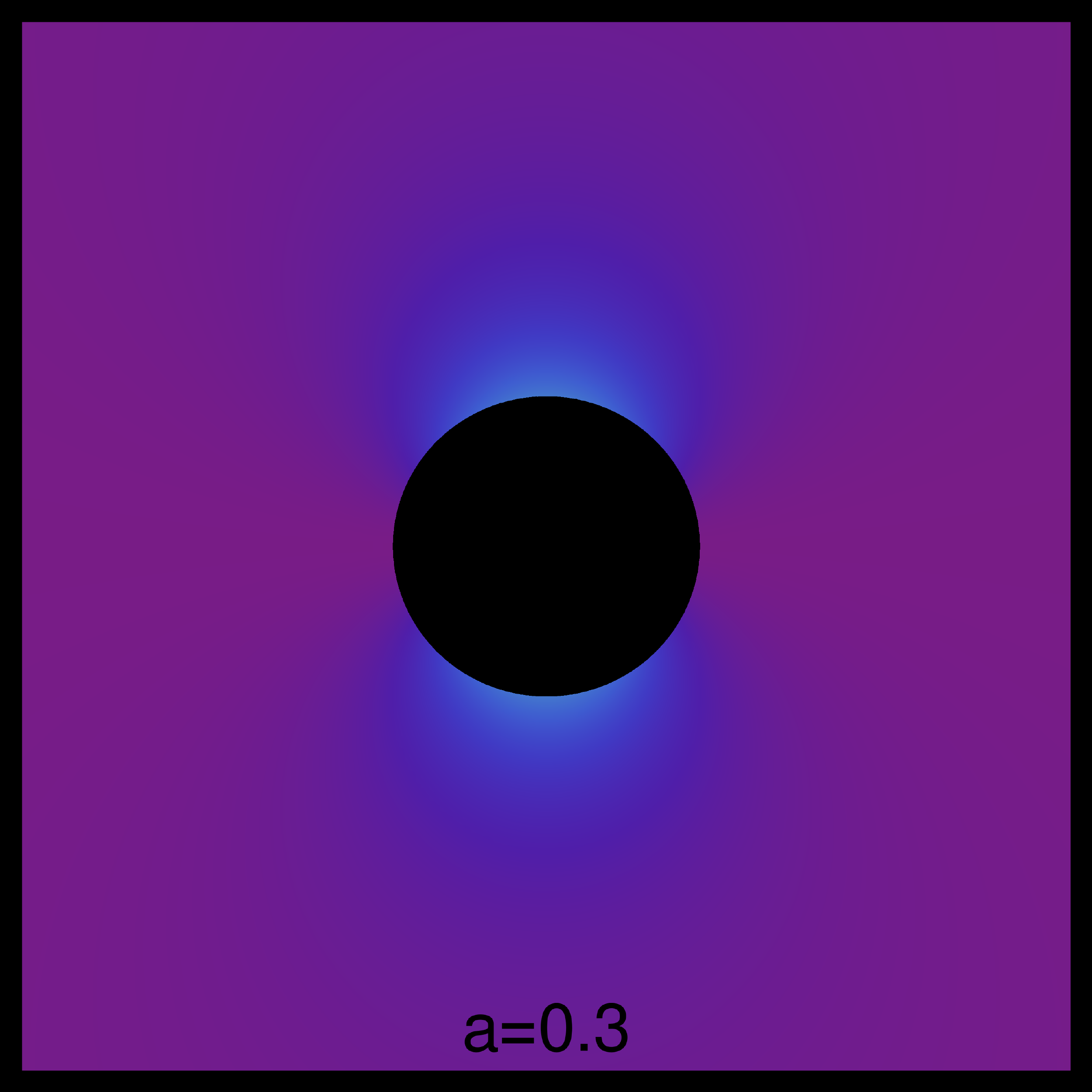} \
	\includegraphics[width=0.035\textwidth]{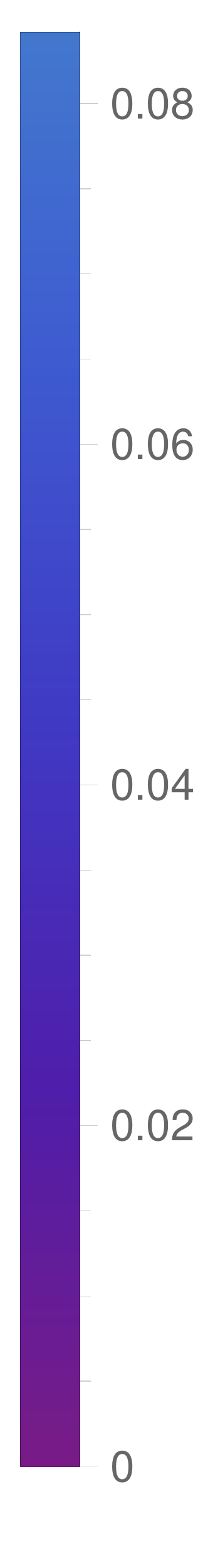} \
	\includegraphics[width=0.27\textwidth]{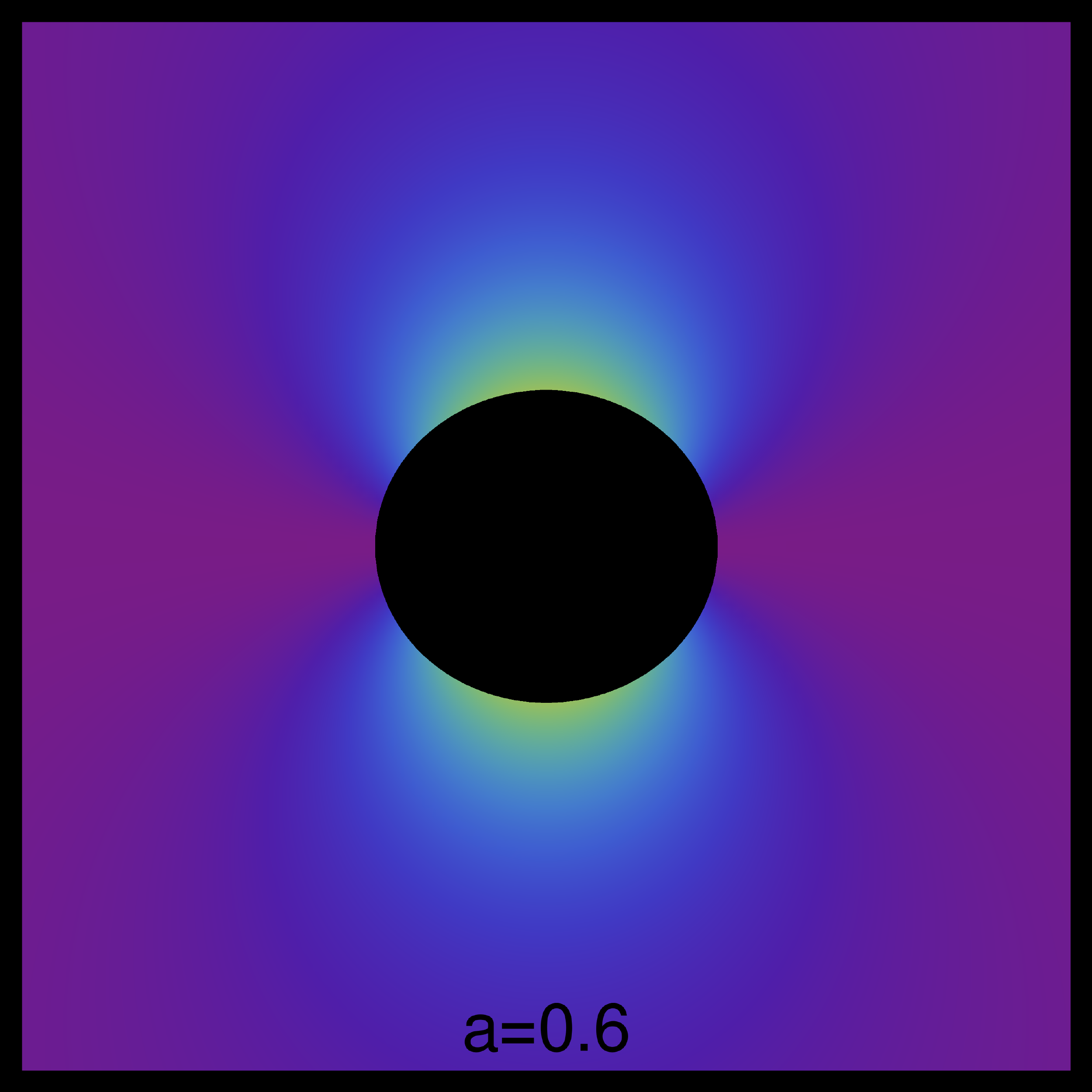} \
	\includegraphics[width=0.035\textwidth]{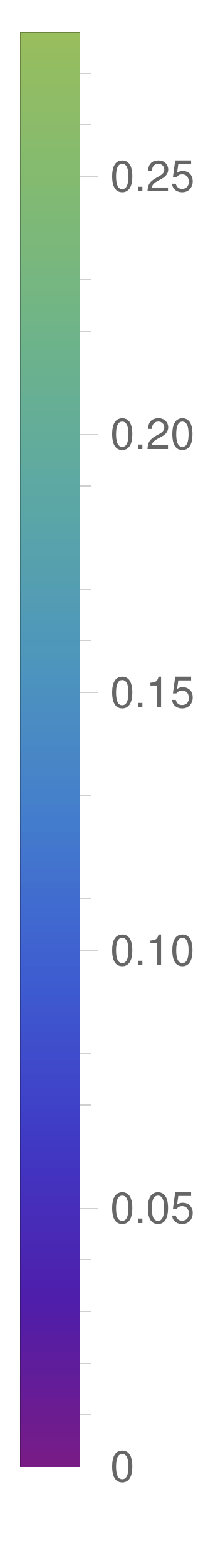} \
	\includegraphics[width=0.27\textwidth]{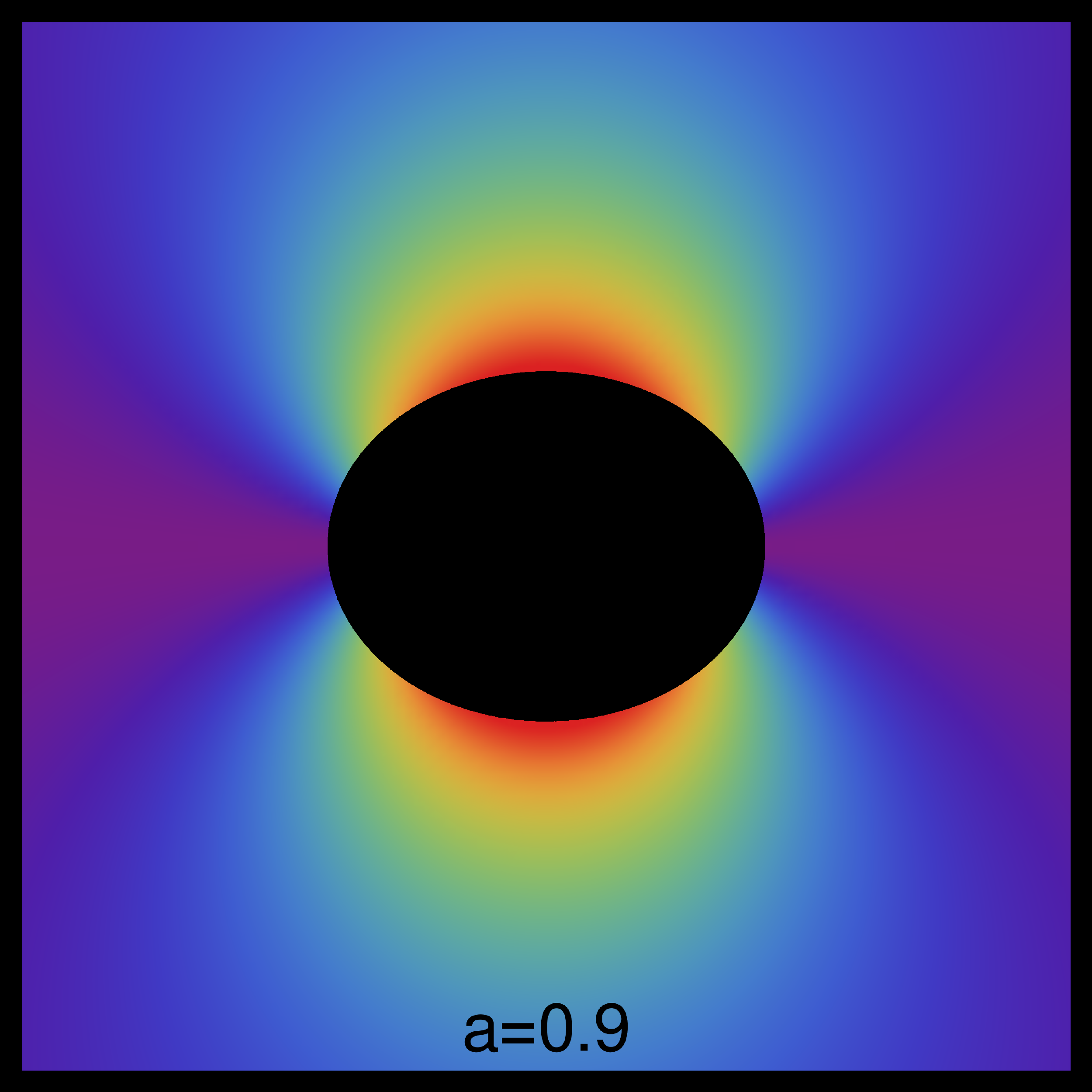} \
	\includegraphics[width=0.031\textwidth]{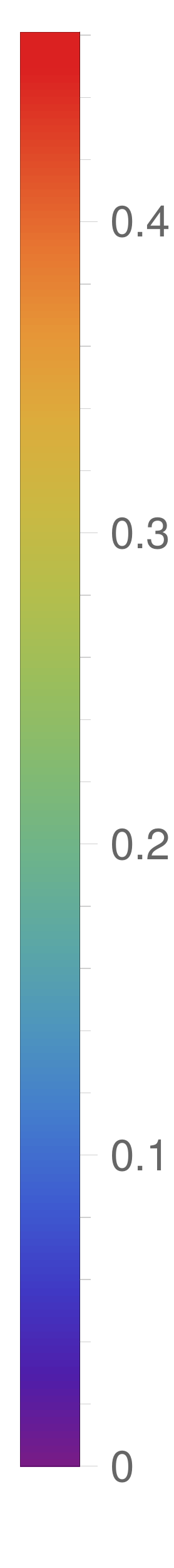} \
	\caption{Anisotropy $\Pi(r,\theta)$ induced around black holes. The equatorial plane remains isotropic, however the anisotropy increases with $a$ and is maximum around the poles, in agreement with Eq.~\eqref{aniso}. We take ${\Lambda/\kappa}=1$.}
	\label{fig2}
\end{figure*}%
whose basic properties we analyze below. First of all, we find that $K_{\mu\nu}$ contains an energy density ${\cal E}$ and pressures ${\cal P}_i$ given by
\begin{eqnarray}
	\label{Ek}
	&&{\cal E}=u_\mu\,u_\nu\,K^{\mu\nu}=0\ ,
	\\
	&&{\cal P}_r=l_\mu\,l_\nu\,K^{\mu\nu}=0\ ,
	\\
	&&\label{Ept}{\cal P}_\theta=n_\mu\,n_\nu\,K^{\mu\nu}={\cal P}_\phi=m_\mu\,m_\nu\,K^{\mu\nu}=\epsilon+p_\theta\ .
\end{eqnarray}
Since ${\cal E}=0$ and ${\cal P}_r\neq\,{\cal P}_\theta$, we conclude that the energy-momentum tensor $K_{\mu\nu}$ is not the source for gradients in the energy density but the cause of anisotropy in the pressures. Second, from Eqs.~\eqref{energyax} and~\eqref{pressuresax}, we see that for the static case (${a=0}$) we have $\epsilon=-p_\theta=\Lambda/\kappa$, and therefore the anisotropic sector vanishes, i.e., $K_{\mu\nu}=0$. Hence, the energy-momentum tensor has the form given in Eq.~\eqref{vac}. However, for the stationary case ($a\neq\,0$) we have a continuous deformation, where the parameter which drives this deformation is precisely the radial coordinate $r$, yielding the following two cases:
\begin{itemize}
	
	\item Far away from the BH ($r>>a$), we have that $\epsilon\sim-p_\theta\sim\Lambda/\kappa$, in consequence $K_{\mu\nu}\sim\,0$, and therefore the energy-momentum tensor is given again by Eq.~\eqref{vac}.
	
	\item Near the BH ($r\sim\,a$) the energy-momentum tensor gets the form in Eq.~\eqref{t3}.
\end{itemize}
We emphasize that the continuous transition in Eq.~\eqref{t2} cannot be arbitrary and, in fact, is subject to the fulfillment of Bianchi identities, which implies that $T_{\mu\nu}$ is covariantly conserved, i.e.,
\begin{equation}
	\label{cons1}
	\nabla_\mu\,T^{\mu\nu}=0\ ,
\end{equation}
which yields
\begin{equation}
	\label{cons2}
	X_\mu\equiv\partial_\mu\epsilon+\nabla_\sigma\,K^{\sigma}_{\,\,\mu}=0\ .
\end{equation}
The expression in Eq.~\eqref{cons2} is quite significant, since it clearly shows an exchange of energy-momentum between the isotropic and anisotropic sector. Indeed, it explains how the latter appears at the expense of the former.

We find that two components of the four-vector $X_\mu$ in Eq.~\eqref{cons2}, namely, $\mu=t,\,\phi$, leads to an identity. However, the radial and tangential components yields, respectively,
\begin{eqnarray}
	\label{er}
	&&\partial_r\epsilon=-\nabla_\sigma\,K^{\sigma}_{\,\,r}=\frac{4a^2\Lambda\,r^3\cos^{2}\theta}{\rho^6}\\
	\label{et}
	&&\partial_\theta{\epsilon}=-\nabla_\sigma\,K^{\sigma}_{\,\,\theta}=\frac{2a^2\Lambda\,r^4\sin{2\theta}}{\rho^6}\ .
\end{eqnarray}
The expressions in Eqs.~\eqref{er} and~\eqref{et} are the hydro-stationary equilibrium equations, and they show in detail what occurs with the vacuum energy density around the BH. From Eq.~\eqref{er} we find that for $\Lambda>0$ and $\theta\neq\,\pi/2$ the energy density $\epsilon$ decreases when we approach to the BH, and therefore we see a flux of energy-momentum from the isotropic sector to the anisotropic one. The opposite happens when $\Lambda<0$. This effect is proportional to $a^2$, and it does not occurs in the equatorial plane, but is maximal in the direction of rotational axis, as we can see in Fig.~\ref{fig1}. Regarding the anisotropy in Eq.~\eqref{aniso}, Fig.~\ref{fig2} shows three different cases.

Finally, we want to emphasize two features of great significance. First of all, the metric~\eqref{newkerrds} does not require the existence of any form of matter-energy  beyond the cosmological constant, since the source of this metric is nothing other than the warped vacuum energy due to the rotating source, as shows by Eq.~\eqref{t2}. Second, we only use general relativity, without resorting to any extended version of it, as the generic one displayed in Eq.~\eqref{ngt}. As we have already pointed out, this seems to contradict the crucial expression in Eq.~\eqref{exchange2}. However, in our case this expression becomes
\begin{equation}
	\label{bianchideformed}
	\partial_\nu\Lambda+\kappa\nabla_\mu\,\cancelto{0}{T}^{\mu}_{\,\,\,\nu}=0\rightarrow\,\,
	{\partial_\mu\epsilon+\nabla_\sigma\,K^{\sigma}_{\,\,\mu}=0}\ ,
\end{equation}
and therefore this apparent contradiction has a simple and fairly straightforward interpretation: both $\Lambda$ and the vacuum $T_{\mu\nu}=0$ in Eq.~\eqref{exchange2} are deformed as 
\begin{eqnarray}
	\label{def1}
	&&\Lambda\rightarrow\,\Lambda(x^\alpha)=\kappa\,\epsilon\\
	\label{def2}
	&&T^{\mu}_{\,\,\,\nu}(=0)\rightarrow\,K^{\mu}_{\,\,\,\nu}\ ,
\end{eqnarray}
leading to Eq.~\eqref{cons2}. Hence, according to the expression in Eq.~\eqref{bianchideformed}, we might say that the cosmological constant creates its own environment with which it exchanges energy.

\section{Conclusions}
\label{con}
In general relativity, the existence of a cosmological constant endowed with local properties, without coexisting with any form of matter-energy, seems to violate Bianchi identities, expressed through the conservation equation~\eqref{exchange2}. However, we find that the above is possible without contravening any principle, in complete agreement with Bianchi identities. 
Our analysis was developed in Kerr-Schild spacetimes, by mean of the revisited Kerr-de Sitter solution~\eqref{newkerrds}. This allows us to identify, as a counterexample, a continuous deformation of the vacuum state in Eq.~\eqref{vac}, such as shown in Eq.~\eqref{t2}, and explicitly explained by Eqs.~\eqref{bianchideformed}-\eqref{def2}. 

Finally, some questions regarding the warped vacuum energy by BHs remain open: for instance, whether the rotational effects shown in Figs.~\ref{fig1} and~\ref{fig2} are only experienced by those fluids satisfying $\epsilon=-p_r$, or perhaps is something intrinsic to extreme gravitational environments. If so, regions of low density and high anisotropy will always appear at the poles. This could give new insight to understand gravitational phenomena along the axis of rotation, such as astrophysical jets. We want to conclude by emphasizing that continuing to insist on an always constant vacuum energy, no matter how extreme the gravitational conditions are, does nothing but hide rotational effects such as those described here.

\subsection*{Acknowledgments}
This research has been partially supported by
ANID FONDECYT grant ${\rm N}^{\rm o}$ 1210041.
%
%

\bibliography{references.bib}
\bibliographystyle{apsrev4-1.bst}
%
%
\end{document}